\documentclass[prl,twocolumn,preprintnumbers]{revtex4-1}%
 
\usepackage{epsf,epsfig}
\usepackage{amssymb,amsmath,amsfonts}
\usepackage{color}
\usepackage{braket}
\usepackage{subcaption}
\newcommand{\bea}{\begin{eqnarray}}
\newcommand{\eea}{\end{eqnarray}}
\newcommand{\bean}{\begin{eqnarray*}}
\newcommand{\eean}{\end{eqnarray*}}

\newcommand{\nn}{\nonumber}

\renewcommand{\dag}{\dagger}

\renewcommand{\H}{\mathcal{H}}

\newcommand{\N}{\mathcal{N}}
\newcommand{\M}{\mathcal{M}}

\newcommand{\om}{\omega}

\newcommand{\be}{\begin{eqnarray}}
\newcommand{\ee}{\end{eqnarray}}

\def\braket#1{\left\langle #1 \right\rangle}

\def\braket#1{\left\langle #1 \right\rangle}

\def\Tr{\mathop{\rm Tr}}

\def\N{{\cal N}}

{}{}
{}{}
\def\vereq#1#2{\lower3pt\vbox{\baselineskip1.5pt \lineskip1.5pt
\ialign{$\m@th#1\hfill##\hfil$\crcr#2\crcr\sim\crcr}}}
\makeatother

\begin{document}
\title{Inertial force, Hawking Temperature and Quantum Statistics}
\author{Yang An}
\affiliation{Zhejiang University, Zhejiang Institute of Modern Physics, 310027, Hangzhou, P.R.China}
\affiliation{Institute for Theoretical Physics, University of Amsterdam, Science Park 904,
1098 XH Amsterdam, The Netherlands}
\email{anyangpeacefulocean@zju.edu.cn}
\begin{abstract}
To explore the mechanism for the entropic force proposal in Entropic Gravity, we propose a specific thermodynamic process for states thermalized in local Hawking Temperature. We find when Casini's version of the Bekenstein bound is saturated, the thermodynamic force derived in the entanglement first law matches the local inertial force for the Schwarzschild solution, except for a negligible statistics-dependent factor.  We argue the gravity viewed by static observers may have observable effects emerged from quantum statistics. The successful detailed calculation in this simple model inspires and is in support of the further development in our following research arXiv:2004.14059. 
% and de-Sitter spacetime.

%To reveal the information content of Schwarzshild spacetime, the complexity growth and entropy for the states are shown to search for geometric duality.

%The derivation is on the way to reveal the mechanics for how microscopic particle generally responds to the gravity from the point view of thermodynamics.
%while in high temperature, we find the formula varies differently. The result implies the Hawking Temperature can lead to the gravitational effect from quantum level, where this simple mechanics may hold for any matter distribution, and suggests potential properties of Hawking Radiation.  
\end{abstract}
\maketitle
\section{I.Introduction}

Three different aspects are combined to consider the inertial force in gravitational field as a thermodynamic dual effect. They are Entropic Gravity \cite{Verlinde:2010hp,Verlinde:2016toy}, Hawking Temperature \cite{Hawking:1974sw}, and research \cite{Marolf:2003sq, Marolf:2004et, Casini:2008cr} on the Bekenstein bound \cite{Bekenstein:1973ur,Bekenstein:1980jp}.

%The Entropic Gravity theory, suggested the gravity as macroscopic phenomenon emergent from information structure of spacetime. Several approaches and proposals have been done based on assumptions, but the answer is far from complete. 
Entropic Gravity is to understand the connection between gravity and thermodynamics involving entropy change. %Entanglement Entropy. 
Several major approaches and proposals have been done \cite{Jacobson:1995ab, Jacobson:2015hqa, Padmanabhan:2009vy,Faulkner:2013ica}, %based on Holographic Principal for ground states, 
while \cite{Carroll:2016lku} examined the major assumptions for the origin of entropy in \cite{Jacobson:1995ab, Jacobson:2015hqa}.

Verlinde's theory \cite{Verlinde:2010hp,Verlinde:2016toy} suggested the gravity as a macroscopic entropic force, emphasizing the existence of an entropy gradient in spacetime responsible for the gravity as the inertial force. Thus the information structure of spacetime influences the gravitational behavior. The open questions to answer are what causes the entropy change
%where does the entropic gradient come from
 and can we form a more specific mechanism to generic situations? % thermodynamic process beyond the near horizon region?

We find clues from historical Hawking Temperature and Casini's work on the Bekenstein bound. % in QFT. %The existence of event horizon is responsible for this only thermodynamic effect. In this very specific situation, which we don't need any knowledge beyond the static mechanics, can we reveal the gravitational force from thermodynamics for the microscopic particles states thermalized in this temperature field?
Noticing Hawking Temperature is position-dependent for the thermal situation seen by different local static observers in the spacetime of asymptotic flat Schwarzschild black holes, we use it to replace the role of Unruh Temperature in common approaches of Entropic Gravity theories. Thus the resulting temperature gradient can %in positions 
vary the entropy and energy of thermal states. % will have entropic gradient % to form specific thermodynamic processes. 
%Ever since the laws of black hole thermodynamics was proposed by Hawking \cite{Bardeen:1973gs}, spacetime was regarded to be emergent from microscopic states which has entropy and temperature. Black holes have Bekenstein-Hawking entropy \cite{Bekenstein:1973ur} and Hawking temperature\cite{Hawking:1974sw}. Unruh was the first to understand the Hawking Temperature as a observer-dependent phenomenon \cite{Unruh:1976db} by considering accelerating observers in Minkowski spacetime, where accelerating observers will find the usual Minkowski vacuum thermalized with particles.

Meanwhile, Casini \cite{Casini:2008cr} proved a precise version of the Bekenstein bound 
%$\Delta S\leq \Delta\braket{K}$
 in quantum field theory which avoided species problem from an argument of non-negativity of relative entropy based on formal research \cite{Marolf:2003sq,Marolf:2004et}. The saturation of the bound was then regarded as a condition of holography \cite{Blanco:2013joa}.  These developments inspired us to relate the subtracted entanglement entropy and energy and bring in the first law of entanglement \cite{Bhattacharya:2012mi} to the improvement of the foundation of Verlinde's theory.%to a specific thermodynamic process seen in the accelerating frame. 

In this paper, we find the existence of the external force measured by accelerating observers as the thermodynamic force turns the situation to be entropic. Thus the entropic mechanism is compatible with Susskind's complexity approach \cite{Susskind:2018tei,Brown:2018kvn,Susskind:2019ddc} to explain the gravitational attraction tendency because they are viewed in different processes. We will show the thermodynamic force is produced exactly from the difference between the modular Hamiltonian $K_1$ of an excited particle state and $K_0$ of the vacuum state
\bea
F_{\mu}=T\nabla_\mu \braket{K-K_1}_1\,.
\eea
where $T=\frac{\kappa}{2\pi V(r)}$ represents the local temperature with the surface gravity $\kappa$ and the redshift factor $V(r)$.%=e^{\phi(r)}$ to the generalized Newton's potential $\phi$.%=\frac{1}{2}\log{(-\xi^\mu\xi_\mu)}$. 

%Based on \cite{Marolf:2004et,Casini:2008cr}, we relate the inertial force to the entropy and energy of the exited states subtracted from thermal vacuum% in free field theory.  
When Casini's version of the Bekenstein bound $\Delta S\leq \frac{\Delta \braket{H}}{T}=S_{\infty}$ is saturated, the thermodynamic expression for $F_{\mu}$ to compare with the local inertial force $\textbf{F}_{inertial}=-ma_\mu$ is
\bea
%F_{\mu}(r)=\frac{1}{1\pm e^{-\omega/T}}\frac{\omega}{T}\frac{\partial T}{\partial r} \delta_\mu^r \,,
F_{\mu}=S_{\infty}\nabla_\mu T= \frac{\nabla_\mu T}{T}\Delta \braket{H}\,,
\label{uniexforce}
\eea
where the subtracted energy $\Delta \braket{H}$ is statistics-dependent and
%where the factor $\frac{1}{1\pm e^{-\omega/T}}$ depends on whether it is Bose-Einstein or Fermi-Dirac statistic and
%\bea
%\frac{\omega}{T} \frac{\partial T}{\partial r}\delta_\mu^r=- {G\M\om\over{r^2(1-\frac{2G\M}{r^2})}}\delta_\mu^r\,,
%\eea
the expression approximates to the local inertial force when applying local Hawking Temperature to $T$.
%where $T(r)=\frac{T_{H}}{V(r)}$ with $V(r)=e^{\phi(r)}$ is the redshift factor to the generalized Newton's potential $\phi$. 

This paper is rather an unperfected story organized to show the coincident derivation of those results in a short cut and then look back to see the conditions and implications of the derivation. In the whole context, we adopt the Natural Unit $c=k=\hbar=1$.
%Thus, the result implies gravitation temperature such as Hawking Temperature also responsible for the inertial force and .

 %The existence of event horizon is responsible for this only thermodynamic effect. In this very specific situation, which we don't need any knowledge beyond the static mechanics, can we reveal the gravitational force from thermodynamics for the microscopic particles states thermalized in this temperature field?

%With the thermal states, which obeys Bose-Einstein statistic or Fermi-Dirac statistic, we know how the entropy and energy of these thermal states changes microscopic states in different temperature, responding to the thermal atmosphere. Even though here is no holography used in the derivation, the response  to meet the require of General Relativity.

\section{II. Thermal Particle States}

\label{setup}

According to the Unruh Effect \cite{Unruh:1976db}, an accelerating observer will see the thermal spectrum of particle states. Even the Minkowski vacuum contains thermal particles to the observer. 
%We draw attention here, the thermal effect comes from the degree of freedom of particle numbers after tracing out the invisible part of spacetime to the accelerating observer. Since the Hawking Temperature is also a observer-dependent effect, we   
%Because of the causal separation, the Hilbert space can be decomposed as a tensor product $\H=\H_{-V}\bigotimes \H_{V}$ with left half space $-V$ and right half $V$ denoting two Rindler wedges. The left half space $-V$ is conventionally taken to be invisible. the Minkowski spacetime is now causal separated into two Rindler wedges.

In \cite{Marolf:2004et,Casini:2008cr}, to prove $\log M$ increasing in species problem will not ruin the Bekenstein bound, 1-particle mixed states of single frequency $\omega$ with $M$ species of fields are considered. In Minkowski spacetime, the mixed states are taken to be $\frac{1}{M}\sum_j|1_j\rangle \langle 1_j| = \frac{1}{M}\sum_j （a_j^{\dag}|0\rangle\langle 0|a_j） $ with the vacuum state $\left|0\right>=\otimes_j |0\rangle_j$ and excited particle added to the right Rindler wedge, where $j$ labels species and $a^\dagger_j$ creates one particle of frequency $\omega$. 

Meanwhile the whole Hilbert space can be decomposed as a tensor product $\H=\H_{-V}\bigotimes \H_{V}$ with respect to two casual separated Rindler Wedges $-V$ and $V$. Tracing over the left Hilbert space $H_{-V}$ which is conventional taken to be invisible, the particle states will follow thermal distribution of Unruh Temperature $T=T_U=\frac{a}{2\pi}$ proportional to the acceleration $a$. 

The reduced density matrixes in \cite{Marolf:2004et}\cite{Casini:2008cr} of the vacuum and 1-particle thermal states with $M$ species of free scalar fields are
\begin{eqnarray}
\rho_V^0&=&  (1-e^{-\omega/T})^M \nn\\ 
&&\sum_{\vec{N}=0}^{\infty}    e^{-\omega {\cal N}/T}
\left|N_1,...,N_M \right>  \left<N_1,...,N_M \right|  \label{den0}\,,\\
\rho_V^1&=&\frac{1}{M}e^{\omega/T}(1-e^{-\omega/T})^{M+1}\nn\\
 &&\sum_{\vec{N}=0}^{\infty}   {\cal N} e^{-\omega {\cal N}/T}\left|N_1,...,N_M \right>  \left<N_1,...,N_M \right|\,, \label{den1}
\end{eqnarray}
 where we adopted a vector notation $\vec{N}=(N_1,...,N_M)$ and the total number operator $\cal N$ satisfying ${\cal N}|N_1,...,N_M \rangle=\sum_{k=1}^M N_k|N_1,...,N_M \rangle$.  One important insight is that they have different modular Hamiltonian.
  
The density matrixes now follow the same thermal distribution as states of $M$ thermal harmonic oscillators with temperature $T$ and single frequency mode $\omega$.

This interesting similarity inspired us to consider the thermal ensembles with the same form of density matrices generally for Bosonic (labeled by $b$) and Fermionic (labeled by $f$) states
\begin{eqnarray}
\rho_0 &=& \frac{1}{Z_0} \sum_{\vec{N}}   e^{-\omega {\cal N}/T}\left|N_1,...,N_M \right>  \left<N_1,...,N_M \right|  \label{den0r}\,,\\
\rho_1&=&\frac{1}{Z_1}\sum_{\vec{N}}  {\cal N} e^{-\omega {\cal N}/T}\left|N_1,...,N_M \right>  \left<N_1,...,N_M \right|\,, \label{den1r}
\end{eqnarray}
where $Z_0$ and $Z_1$ are normalization factors satisfying $\Tr{\rho_i}=1$. However, the basis are different for the Fermionic states with $N_{j}\in\{0,1\}$ due to Fermi-Dirac statistic. %Later we will put the situation more general considering $T=T(r)$ for thermodynamic process in curved spacetime.
 
The expectation value of the number operator $\N$ of the Bosonic thermal vacuum would be
\bea
\braket{\N}^{b}_{0}&=&\frac{M}{e^{\omega/T}-1}\,,
\eea
which agrees with Bose-Einstein statistic, while that of the thermal Fermionic vacuum would be
\bea
\braket{\N}^f_{0}&=&\frac{M}{e^{\omega/T}+1}\,,
\eea
which is what Fermi-Dirac statistic would tell. 

Difference from statistics will appear when the states are confined in subsystem as thermal states. %Will the statistic-dependence remain?
In the following context, we make our calculation using the expression of $\rho^i_V$ for the Bosonic ensembles, then generalize the result to Fermionic ensembles.
\subsection{II.1 Observer-Dependent Energy and Entropy}

The Hamiltonian of the quantum oscillators is 
\bea
H &=&\omega \N \,,
\label{hamiltonian}
\eea
where $\N$ is the number operator counting the total number of particles with single frequency $\omega$ mode. 

The energy of any state $\rho$ is calculated by counting the expectation of the number operator $\N$

\bea
%E(\rho)=
\braket{H}_{\rho}&=&\omega \braket{\N}_{\rho}=\omega\Tr{\rho \N}\,,
\eea
%where we need to take the expectation value of the total number of the particles of the states.
%For a state $\rho$,
and the Von Neumann entropy is
\bea
S(\rho)=-\Tr(\rho \log{\rho})~,~~~
\eea
which is also the entanglement entropy for reduced density matrixes.

Because of Unruh Effect,
% when regarded as thermal harmonic oscillators,
we notice  $\rho_V^i$  corresponding to the single mode in consideration is connected to Bose-Einstein statistics. The energy is
\bea
\braket{H}_0&=&\frac{M\omega}{e^{\omega/T}-1}\,,\\
\braket{H}_1&=&\frac{\omega}{1-e^{-\omega/T}}+\frac{M\omega}{e^{\omega/T}-1} \nn\\
&=& \frac{\omega}{1-e^{-\omega/T}}+\braket{H}_0\,,
\eea
%which are statistic-dependent.
and the entanglement entropy of the vacuum is
\begin{equation}
S(\rho_V^0)=M\left(\frac{w/T}{e^{\omega/T}-1}-\log\left(1-e^{-\omega/T}\right)\right)\,,
\label{srv0}
\end{equation}
which is the entropy of a thermal ensemble of $M$ independent oscillators,
% It also follows that 
%\begin{equation}
%S(\rho_V^1)=-\sum_{\vec{N}=\vec{0}}^\infty \frac{1}{M}e^{\omega/T}(1-e^{-\omega/T})^{M+1} {\cal N} e^{-\omega/T {\cal N}} \log\left(\frac{1}{M}e^{\omega/T}(1-e^{-\omega/T})^{M+1} {\cal N} e^{-\omega/T {\cal N}} \right)\, , \label{tt}
%\end{equation}
while the entropy of $\rho_V^1$ is
\bea 
S(\rho_V^1)&=&S(\rho_V^0)+\log(M)-\log(e^{\omega/T}-1)+\frac{\omega/T}{1-e^{-\omega/T}}\nn\\
&-&\sum_{\vec{N}=\vec{0}}^\infty\rho_V^1 \log {\cal N}\,,
\label{S1logNf}
\eea
where the last term $\sum_{\vec{N}=\vec{0}}^\infty\rho_V^1 \log {\cal N}$ is just $\braket{\log \N}_1$.
 
It is no hard to generalize the results to be Fermionic by changing statistics-dependent factors. Noticing the statistical differences already appear in energy and entropy for different quantum statistics and also for different species number $M$, will the negligible effect remain for gravity in an entropic mechanism? %Though the effect may not be obviously observable when $T\rightarrow 0$, will the effect remain for gravity in an entropic mechanism? %Also, we put the question here if the statistical dependence remains when consider the gravity as an entropic force.

\subsection{II.2 Entropy Bound in Large Species Number Limit}

When $M$ is very large so $Me^{-\om/T}\gg 1$, we can follow \cite{Marolf:2004et} for a mean-field expansion to evaluate the last term in (\ref{S1logNf}):
\bea
\label{lM}
\braket{\log{\N}}_1=\log{\braket{\N }_0}+O(\frac{1}{Me^{-\omega/T}})\,,
\eea
%For thermal 1-particle states of Fermionic fields, if the density matrix is of the form of (\ref{den1r}), we guess that the mean-field expansion will lead to
%\bea
%\braket{\log \N_f}_1=\log{\braket{\N_f }_0}+O(\frac{1}{Me^{-\omega/T}})=\log{\frac{M}{e^{\omega/T}+1}}+O(\frac{1}{Me^{-\omega/T}})
%\eea
and we will get
\bea
S(\rho^1_V)-S(\rho^0_V)\leq \frac{w/T}{1- e^{-\omega/T}}\,.
\eea
In analogue, for thermal states, we will get
\bea
S(\rho_1)-S(\rho_0)\leq \frac{w/T}{1\pm e^{-\omega/T}}\,,
\eea 
where $+$ for Fermionic states while $-$ for Bosonic states.
Generally, the bound is hold because relative entropy  $S(\rho_1||\rho_0)$ is non-negative
\bea
 S(\rho_1||\rho_0)\geq0\,,
 \label{CBekenbound}
%\frac{w/T}{1\pm e^{-\omega/T}}-(S(\rho_1)-S(\rho_0))
\eea
which is equivalent to Casini's version of the Bekenstein bound \cite{Casini:2008cr}
\bea
S(\rho_1)-S(\rho_0)\leq (\braket{H}_1-\braket{H}_0)/T\,,
\eea
where $H/T$ is also the modular Hamiltonian $K_0$ of vacuum.
When $M\rightarrow\infty $, the bound $S_{\infty}$ is saturated
\bea
S_{\infty}=\lim_{M\rightarrow\infty} \left(S(\rho_1)-S(\rho_0)\right)=\frac{w/T}{1\pm e^{-\omega/T}}\,,
\label{casibound}
\eea
and the relative entropy $S(\rho_1||\rho_0)$ becomes zero, which means the 1-particle state is highly mixed and indistinguishable from the vacuum. The condition is closely related to holography \cite{Blanco:2013joa}, so it provides hints for a holographic explanation later developed in \cite{An:2020ncr}.  %The connection to the black hole holography will be find in approaching research to show that the bound is the change of black hole entropy when absorbing the excited state \cite{}. % and the states are maximal-entangled.

%The bound is also statistical different for Bosonic and Fermionic states.
%Besides, since any 1-particle pure state $|1\rangle_i = a_i^{\dag}|0\rangle $ with M species vacuum $\left|0\right>=\otimes_i |0\rangle_i$ is equivalent to the state with $M=1$, the reduced density matrixes for the pure excited states have different subtracted entropy and the bound is far from statured.
%We will see the bound plays an important role in the expression of inertial force.
\section{III. Thermodynamics}
We consider distributions of states change with infinitesimal variation with respect to the parameters $\om/T$ first, so the variation of function $g(\om/T)$ will vary as
\bea
\delta g(\om/T)=\frac{d g(\om/T)}{d{(\om/T})}\delta(\om/T)\,.
\eea
Then we will distinguish the variation of frequency $\om$ and temperature $T$ separately, they are relevant to different thermodynamic processes.

\subsection{III.1 Work term from two First Laws}
%The thermodynamic first law is well known as the conservation law of energy while the entanglement first law is formed for entanglement entropy.
Here we compare the thermodynamic first law and the entanglement first law, to locate the work term from their difference. 

First we write down the thermodynamic first law
\bea
	\delta Q+\delta W=\delta E\,.
	\label{thermofirst}
\eea

We checked the following equalities are satisfied
\bea
T\delta S(\rho_0)&=&\delta \braket{H}_0\,,
\label{entfirst0}
\\
T\delta S(\rho_1)&=&-T\delta \braket{\log{N}}_1+\delta\braket{H}_1\,,
\label{entfirst1}
\eea
which indeed agrees the entanglement first law $dS=d\braket{K}$
where the modular Hamiltonian $K$ is defined from a reduced density matrix $\rho=\frac{e^{-K}}{\Tr{e^{-K}}}$. We notice $K_0-K_1=\log N$ and $K_0=H/T$. 

One can compare (\ref{entfirst0}) and (\ref{entfirst1}) with the thermodynamic first law (\ref{thermofirst}) to get
\bea
\delta W+T\delta S(\rho_1)=\delta\braket{H}_1\,.
\label{heat}
\eea

%Substract the vacuum from the 1-particle states, a good guessing is
We can relate the work term $\delta W$ for 1-particle states to the term $T\delta \braket{\log{\N}}_1$, where $\log{\N}$ is just the difference of two modular Hamiltonians
\bea
\delta W =T\delta \braket{\log{\N}}_1=T\delta \braket{K_0-K_1}_1\,.
\label{exwork}
\eea

%The external work indeed is the difference of the modular Hamiltonians of excited stats and vacuum.
This is the first result we have got.

\subsection{III.2 Fixed Frequency Process}
\begin{figure}[h]
\centering
\includegraphics[scale=0.6]{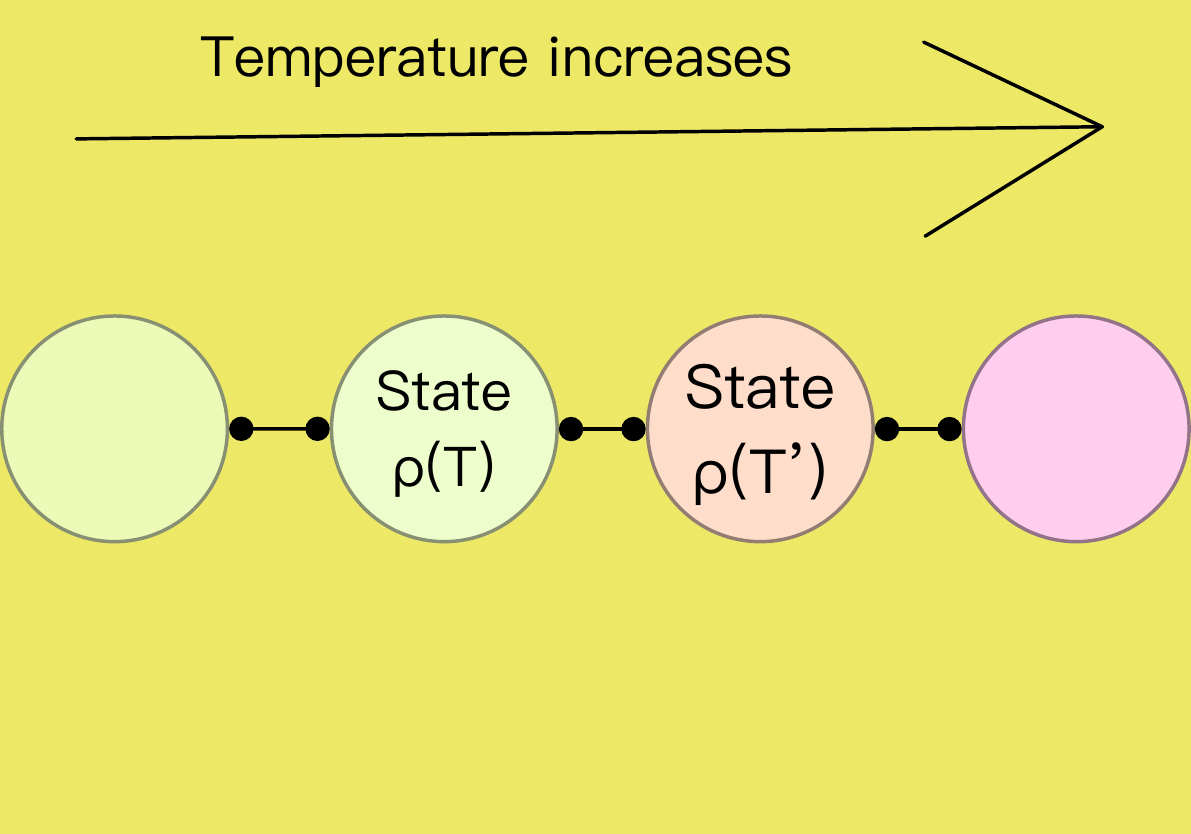}
\caption{A test exited state $\rho(T)$ is hold by an external force $F_{\mu}$ in a static thermal atmosphere of position-dependent temperature. $F_{\mu}$ can be calculated by varying the states to be $\rho(T')$ nearby. % in a quasi-static process that doesn't change the momentum $dp=0$.
%with $dp=0$. 
This picture is in a triumph to relate the change of thermal states, to the phenomenon of gravity.}
%The falling tendency can be explained as the states want to "jump" to the nearby state with higher temperature, to gain more energy and entropy.
\label{Ttensor}
\end{figure}
A thermal reversible progress is an ideal quasi-static process that changes the ensemble while keeping it always in equilibrium with the outside heat bath.%, allowing to subtract the maximal external work. 

%We can change the thermal ensembles in two way, first is changing the $\omega$, the other is heating them to change $T$. The first one will give external force the energy of the particle $\Delta W\approx \Delta \omega$, which agrees with what a Minkowski observer would see how the energy changes, now we focus the second one.

We propose a specific thermodynamic process in this spirit when we have a temperature gradient $\nabla_{\mu}T$ as shown in FIG.\ref{Ttensor}. We assume exact exterior influence is done to the states so that the frequency $\om$ won't change, while the particle state gains no momentum change during the process: $dp=0$.

If the temperature of the states depends on position parameter $r$ while its frequency keeps fixed, we have
\bea
\nabla_{\mu}g(\om/T)= \partial_rg(\om/T)=\delta^r_\mu\frac{\partial T}{\partial r}\partial_Tg(\om/T).
\eea
The covariant derivative of a scalar function depends on how the temperature varies with position viewed by the observer moving along with the states.
%Thus each single mode keeps  during the process. 

%because
%\bea
%\omega^2=m_0^2+p^2
%\eea
%for any Bosonic and Fermionic fields, where $m_0$ is the rest mass of the particle measured by observers at infinity of asymptotic flat. 

\subsection{III.3 Derivation of Inertial Force}

The local inertial force should be opposite to the external force $\textbf{F}_{inertial}=-ma_\mu$. From (\ref{exwork}) we can derive the thermodynamic force as %using $F_{\mu}(r) dr =\Delta W$:
\bea
F_{\mu}(r)=T\nabla_\mu \braket{log \N}_1\,.
\label{extforcedef}
\eea
This expression depends on species number $M$. However, when the bound is saturated (\ref{casibound}) at $M\rightarrow\infty$, the general expression for $F_{\mu}(r)$ is 
\bea
F_{\mu}(r)
&=&
	-T\frac{\partial T}{\partial r} \partial_T \left(\log \left(e^{\omega/T}-1\right)\right) \delta^r_\mu
	\nn\\
&=& 
	-T\frac{\partial T}{\partial r} (-\frac{\omega}{T^2})\frac{e^{\omega/T}}{e^{\omega/T}-1} \delta^r_\mu
	\nn\\
&=& \frac{\omega}{T} \frac{\partial T}{\partial r} \frac{1}{1-e^{-\omega/T}} \delta^r_\mu\,.
\label{exfactor}
\eea
Since $e^{-\omega/T}\approx 0$ in low temperature limit $T\rightarrow 0$, we can get an approximation
\bea
F_{\mu}(r) &\approx &
 \frac{\omega}{T} \frac{\partial T}{\partial r} \delta^r_\mu\,.
 \label{lowexforce}
\eea

Now we put the situation in a fixed curved spacetime background. For a asymptotical flat Schwarzschild black hole of mass $\M$, with the metric $ds^2=-f(r)dt^2+{1\over f(r)}dr^2+{r^2}d\Omega_{2}$ and $f(r)=g_{tt}=\frac{1}{g_{rr}}=1-\frac{2G\M}{r}$, the Hawking radiation has the Hawking Temperature $T_H$ at infinity. For a static observer at $r$ with four-velocity $U_\mu=(1,0,0,0)$, he will observe a local temperature blue-shift from Hawking Temperature $T_H$ that follows Tolman law (see \cite{Wald:1999vt})
\bea
 T(r)={T_H\over {V(r)}}={\kappa\over {2\pi V(r)}}\,,
 \label{hawkingT}
\eea
where $\kappa=\frac{1}{4GM}$ is the surface gravity of the event horizon and $V(r)=\sqrt{1-\frac{2G\M}{r}}$ is the redshift factor.

This is a natural candidate for the origin for the temperature gradient. Applying the local Hawking temperature (\ref{hawkingT}) at position $r$ to the right side of (\ref{lowexforce}), we get
\bea
\frac{\omega}{T} \frac{\partial T}{\partial r}  ={\om V(r)}\partial_{r}\frac{1}{V(r)}=- {G\M\om\over{r^2(1-\frac{2G\M}{r})}}\,,%=- {G\M\om\over{r^2(1-\frac{2G\M}{r})}} \delta^r_\mu\,,
\label{exforce}
\eea
which matches the local inertial force for the Schwarzschild solution
\bea
\textbf F_{{inertial}}={\om}\partial_{r}\phi=-{G\M\om\over{r^2(1-\frac{2G\M}{r})}}\delta^r_\mu\,,
\eea
rather than external force $ma_\mu$, where $V(r)=e^{\phi(r)}$ and $\phi(r)$ is the generalized Newton's potential. At the same time, the entropic gradient during the process is indeed in the opposite direction to that of Verlinde's original thought, as later we will show in \cite{An:2020ncr}. 

The derivation also works for states following Fermi-Dirac statistics by replacing the factor in (\ref{exfactor}) to $\frac{1}{1+e^{-\omega/T}}$. We will see how negligible the statistics-dependent factor $\frac{1}{1\pm e^{-\omega/T}}$ is. The approximation in (\ref{exforce}) is indeed very precise for low temperature. 

\subsection{III.4 Free Falling Process}

If we change $\om$ without changing temperature $T$, the work term will be
\bea
\delta W_{\om} =T \partial_{\om}\braket{\log{\N}}_1 d\om &=&\partial_{\om/T}\braket{\log{\N}}_1 d\om\,.
\eea
Compare with the thermal reversible process which keeps $\om$ constant
\bea
\delta W_{r}=T \partial_{r}\braket{\log{\N}}_1 dr &=&-\frac{\om}{T} \frac{\partial{T}}{\partial r}\partial_{\om/T}\braket{\log{\N}}_1 dr \,,
\eea
if there is no net external work
\bea
\delta W_{\om}+\delta W_{r}=0\,,
\eea
we will get
\bea
d\om &=&\frac{\om}{T} \frac{\partial{T}}{\partial r} d r=-{G\M\om\over{r^2(1-\frac{2G\M}{r})}} d r\,.
\eea
The part $\partial_{\om/T}\braket{\log{\N}}_1$  no longer exists! The result agrees with the gravitational redshift effect in General Relativity but we get it from a  virtual thermal process, no matter whether the Casini's version of the Bekenstein bound is saturated or not.%. And this result holds for any specie number $M$.

Here we have $F_\mu=0$ and no entropy change. Actually, it is the external force that maintains the static orbit, and causes the thermodynamics.

%Now we get the version of thermal process dual to free falling. We can imagine it from the point of view of infinite static observers who sit in a chain to observe the states evolving in $T(r)$, but leave no external work to the states in total. 

\section{IV. Looking back to the Result}

\paragraph{The Principle of Equivalence} 

The above results lead us to relate the gravitational mass to the frequency
\bea
m=\om\,.
\eea

The mechanism inspired from the simple model should work universally as gravity. If each frequency mode $\om_j$ of a more complex state doesn't cross into each other during the fixed frequency process, the mode behaves as independent 1-particle single mode.
%we have infinite number of harmonic oscillators, for any number of particles with different modes,
%\bea
%\sum_i (a^{\dag}_{i,\omega_j})^{n_j}|0\rangle \langle 0|(a_{i,\om_j})^{n_j}
%\eea
%the total energy and entropy would be by sum out all particles. 
%Complex matter may behave like independent 1-particle states. So our results from simple model of 1-particle with single modes are more general, even if the excited states are pure. 
The gravitational effect then follows the superposition principle,
%\bea
%m=\sum_j \omega_j\cdot n_j\,,
%\eea
so our results from the simple model of 1-particle states of single mode can be generalized into general formula for the external force, regardless of the detail of the states.

However, the new thing is the factor $f=\frac{1}{1\pm e^{-\omega/T}}$.  When $M\rightarrow\infty$,  $F_{\mu}$ in (\ref{exfactor}) depends on the subtracted energy $\Delta \braket{H}$ measured by the static observer. This is consistent with the Principle of Equivalence where gravity gets response to any form of energy, and origins from the same mechanism viewed by accelerating observers. We will give it a general proof in \cite{An:2020ncr}.%even to $\Delta \braket{H}$ in the non-inertial frame.% to get correct response to any form of energy, even it is an observer-dependent energy. 

%Notice the Newtonian acceleration is the first order correction in special approach, however, 

\paragraph{Statistics Behavior}

\begin{figure}[h]
\centering
\includegraphics[scale=0.7]{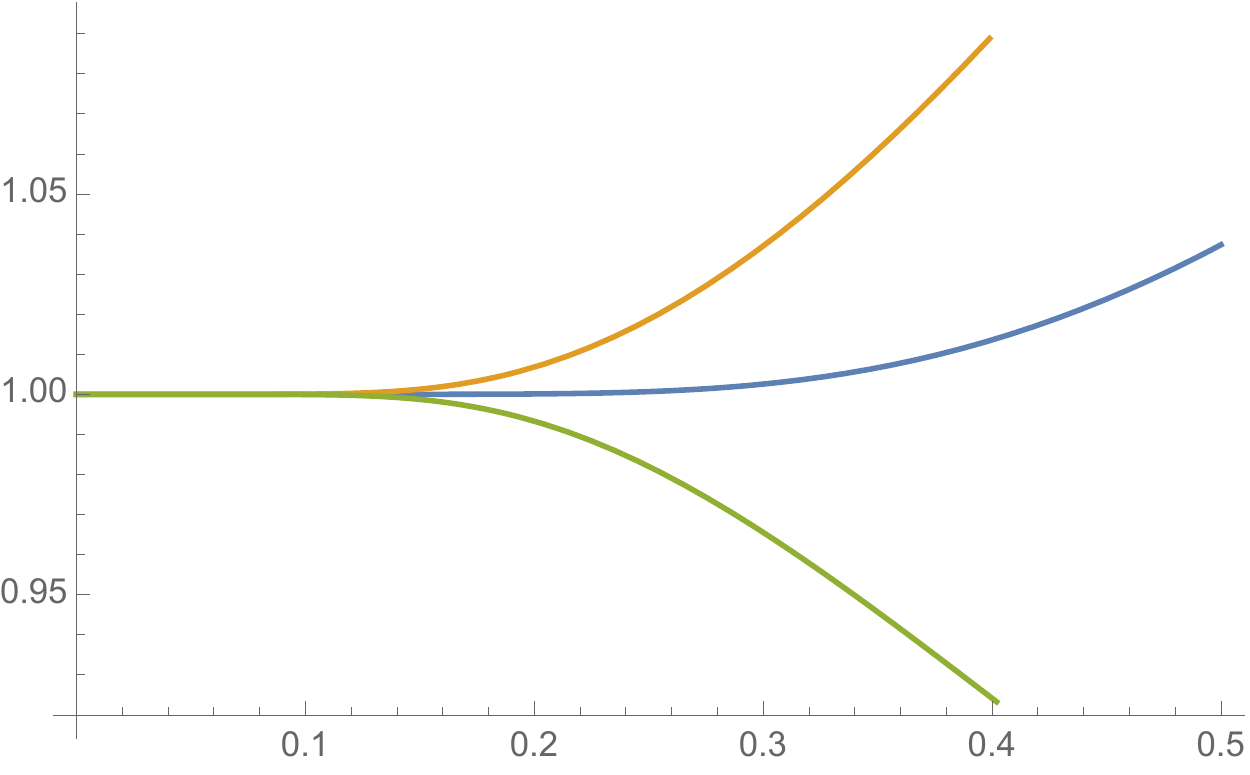}
\caption{The error from $\epsilon=1-f$  is less than $\pm 5 \%$ at ratio
%$\frac{\hbar \om}{kT}=0.3$
$\frac{T}{\om}\leq 0.3$, where  orange line is for Bosonic states with and green line is for Fermionic states. The difference can be balanced for Supersymmetric mixed states, see the blue line.}
\label{omT}
\end{figure}

Does gravity seen by the observers in the non-inertial frames as the external force do depend statistics and how much the states are mixed? First we notice that for Unruh Effect, by counting the number of particles, 
%the Rindler observers already has different definition of Energy and Entropy, though the quantities may not be conserved.
the nearly negligible statistics-dependent factor $f=\frac{1}{1\pm e^{-\omega/T}}$ in inertial force already occurs in energy and entropy. As shown in FIG.\ref{omT}, the staticistics-dependent factors is very close to 1 when the ratio $T/\om$ is small. Besides, when building up a Supersymmetric mixed state to simulate the classic matter mixed with Bosons and Fermions by replacing the basis $|N_j\rangle\langle N_j|\mapsto|N_{f,j},N_{b,j}\rangle\langle N_{f,j},N_{b,j}|$ with $N_{j}=N_{f,j}+N_{b,j}$, the difference can be balanced between two kinds of statistics. 

%When $M\mapsto \infty$, we have seen the surface gravity $\kappa$ ends up with no effect in (\ref{exforce}) after the low-temperature approximation. What matters is not the but how temperature changes with space. 
We do a numerical calculation for a case of a black hole of the Sun mass $\M=\M_{\bigodot}$ where Hawking Temperature is $T_H\approx 6.170\times 10^{-8}$K and using small frequency $\om=10^{14}$Hz of a usual photon. The position $r$ with error $\epsilon=5\%$ is still very close to the event horizon at radial $r_s$: $r(\epsilon=5\%)-r_s\approx 0$ to the accuracy of $10^{-23} r_s$. The difference from statistics may be hard to test until the acceleration is very large.

%The surface gravity $\kappa$ ends up with no effect in (\ref{exforce}) after the low-temperature approximation. What matters is that the temperature changes with redshift factor.

However, the effect from different species number $M$ may be still testable. Since the inertial force for accelerating observers almost matches with Schwarzschild solution only when the Casini's version of the Bekenstein bound is saturated, observation of gravity as the external force in the view of static observers % inertial force
 may differ between quantum pure states ($M=1$) and mixed states.% for classical matter.

%not only because of the statistics-dependent factor., as the consequence of Unruh Effect and Hawking Effect.

\section{V.Implications and Discussion}
The coincident derivation has some further possible implications. We discuss three major implications towards our results.

Firstly, we now have a dual thermodynamic explanation of the inertial force indirectly from  $\textbf F_{{inertial}}=F_\mu$. When Casini's version of the Bekenstein bound is saturated, we notice the free energy $U_i=\braket{H}_i-TS(\rho_i)$ of the 1-particle state and vacuum are the same. Thus from 
\bea
\delta U_i=-S(\rho_i)\delta T+\delta W\,,
\eea
where $\delta W=0$ for $\rho_0$ and $\delta W=|F_\mu| |\delta r|$ for $\rho_1$, $F_{\mu}(r)$ can be generalized into the formula (\ref{uniexforce}) made of the entropy bound $S_{\infty}$ and temperature gradient because
\bea
F_{\mu}(r)=\lim_{M\rightarrow \infty} (S(\rho_1)-S(\rho_0))\nabla_{\mu} T\,,
\label{FSdT}
\eea
rather than of entropic force expression $F_a=T\nabla_a S$ for the inertial force in \cite{Verlinde:2010hp}.  That is because when considering statistics, the heat flux $\delta Q=T\delta S$ not only transforms into external work $\delta W$, but also into internal energy $\braket{H}$. We will show that $F_a=T\nabla_a S$ can be approximated from (\ref{FSdT}) when omitting the statistical effect in future research \cite{An:2020ncr}. 

% in AdS/CFT correspondence to
Secondly, the role of large species number limit resembles the large N limit to form a classical limit. But even if there are not so many different species of fields, the bound may be easily achieved for classical matter distribution  as highly mixed states. For example, by referring to the directions of momentum $\vec{k}$ to replace the role of different species of field, there are infinite eigenstates with the same frequency $\omega$
\bea
a_j^\dagger \mapsto a^\dagger_{\vec{k}_j}\,.
%|N_{\om,i}\rangle \mapsto|N_{\om,\vec{p}_i}\rangle\,.
\eea
So far without any thing holographic, the mechanism works out fine. But the condition of large species number limit connecting with vanishing of relative entropy promises a holographic origin of gravitational attraction for the entropic mechanism.

Thirdly, we have appointed a new role of Hawking Temperature to replace that of Unruh Temperature in Entropic Gravity theories. The free-field model in the temperature gradient is representative to explain the origin of the entropic gradient for gravity. That is because the gravity should response to any form of energy and obeys a universal mechanisms. The entropic mechanism doesn't depend on the detail form of states.

\begin{figure}[h]
\centering
\includegraphics[scale=0.5]{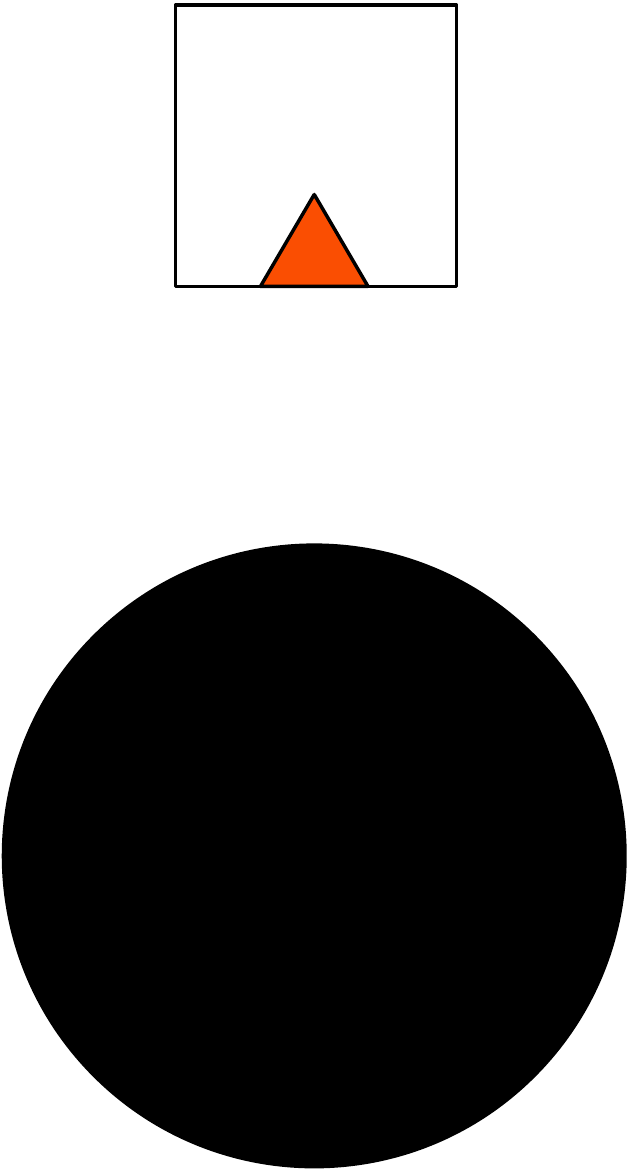}
\caption{A static box in the spacetime a black hole. The triangle inside the box also feels gravity of the black hole. If the temperature gradient is the reason for gravity, can the temperature be prevented by the box?}
\label{box}
\end{figure}

%Though we only focus on the free field theory in the absence of holographic principal, the simple situation is powerfully representative for two reasons. First, the  gravity should response to any form of energy, and the second is that even to a bounded states or interactive system, the energy of free field theory forms major part of total energy.
We notice that Hawking Temperature is also an observer-dependent effect like Unruh Temperature, and the entropy and energy come from the microscopic degrees of freedom of particle numbers. Thus the above mechanism relys on those degrees of freedom independently. This kind of temperatures can not be insulated, as a thought experiment FIG.\ref{box} showing the logic.

%Since it can have macroscopic Temperature higher than Hawking Temperature, the gravitational temperature should rely on independent degrees of freedom.

However, the question comes if general matter distributions such as the Sun and stars should also result in the same temperature as Hawking Effect?

Since in our situation we have argued the mechanism works for black hole and the inertial force is same in the Schwarzchild solution. We have reasons to infer the same mechanism also works for general matter distribution: the general matters also thermalizes the particle states with an equivalent gravitational temperature field $T(r)$. It is possible that the entanglement structure of spacetime for general matter distribution may also exist in the same way as black holes.

In a summary of this paper, thermodynamic processes were built from the point of view of accelerating observers, where we already know the existence of temperature, so degrees of freedoms same in Unruh Effect which are observer-dependent appeared to explain the familiar gravitational results in GR but saw the influence of quantum statistics.

%According to the local observation, he can infer where time Killing Vector vanish to form a Killing Horizon from local experiments of curvature but can not tell if the horizon truly exist according to if there is a local temperature. Same effect happens for the inertial force with local temperature caused by the same mass of the black hole. 

%include surface gravity, can the thermodynamic equation lead us to more general result of the General Relativity or even Quantum Gravity? And does the temperature exists even when no Killing Horizon is there? 

%\paragraph{Equivalent Principal}

%Temperature different degree of freedom.
%
%Distinguish Thermal temperature and gravitational temperature.
%
%The little difference for degree of freedom of Bose and Fermi is also presented but has almost no effect to the inertial force at the region where general relativity holds. 
%
%
%To remind what we already know 

\section{Acknowledge}
Yang An would like to thank: Adolfo Toloza, Peng Cheng, Damian A. Galante, Guilherme L. Pimentel in University of Amsterdam for useful discussions; Mingxing Luo, Bo Feng, Guohuai Zhu and Kai Wang in Zhejiang Institute of Modern Physics for years of supports; speakers in 2018 Arnold Sommerfeld School such as Raphael Bousso, Gia Dvali, Larus Thorlacius, Kyriakos Papadodimas, Samir Mathur for interesting lectures and answering my questions. 

The special gratitude is to Erik Verlinde and Jan de Boer for instructions and invitation to visit University of Amsterdam. This work is supported by visiting PhD scholarship of Zhejiang University, and the National Natural Science Foundation of China (NSFC) with Grant No.11135006, No.11125523 and No.11575156.
\bibliographystyle{JHEP}
\bibliography{Pthermoinerforce}

\end{document}